\documentclass[12pt,a4paper,final]{iopart}
\usepackage{iopams}

\usepackage{graphicx}
\usepackage{dcolumn}
\usepackage{float}
\usepackage{iopams}
\usepackage{multirow}
\usepackage[breaklinks=true,colorlinks=true,linkcolor=blue,urlcolor=blue,citecolor=blue]{hyperref}

\begin{document}

\title{Electron-boson spectral density of LiFeAs obtained from optical data}

\author{J. Hwang$^1$, J. P. Carbotte$^{2,3}$, B. H. Min$^4$, Y. S. Kwon$^4$ and T. Timusk$^{2,3}$}
\address{$^1$Department of Physics, Sungkyunkwan University, Suwon, Gyeonggi-do 440-746, Republic of Korea \\$^2$Department of Physics and Astronomy, McMaster University, Hamilton, ON L8S 4M1, Canada \\$^3$The Canadian Institute for Advanced Research, Toronto, ON M5G 1Z8 Canada \\$^4$Department of Emerging Materials Science, Daegu Gyeongbuk Institute of Science and Technology (DGIST), Daegu 711-873, Republic of Korea}

\ead{jungseek@skku.edu}

\date{\today}

\begin{abstract}
We analyze existing optical data in the superconducting state of LiFeAs at $T =$ 4 K, to recover its electron-boson spectral density. A maximum entropy technique is employed to extract the spectral density $I^2\chi(\omega)$ from the optical scattering rate. Care is taken to properly account for elastic impurity scattering which can importantly affect the optics in an $s$-wave superconductor, but does not eliminate the boson structure. We find a robust peak in $I^2\chi(\omega)$ centered about $\Omega_R \cong$ 8.0 meV or 5.3 $k_B T_c$ (with $T_c =$ 17.6 K). Its position in energy agrees well with a similar structure seen in scanning tunneling spectroscopy (STS). There is also a peak in the inelastic neutron scattering (INS) data at this same energy. This peak is found to persist in the normal state at $T =$ 23 K. There is evidence that the superconducting gap is anisotropic as was also found in low temperature angular resolved photoemission (ARPES) data.

\end{abstract}


\vspace{2pc}
Article preparation, IOP journals
\submitto{\JPCM}

\section{Introduction}

The fluctuation spectrum of inelastic scattering of charge carriers in metallic systems is an essential element in understanding their superconductivity. In conventional $s$-wave superconductors the important mechanism is the electron-phonon interaction\cite{carbotte:1990}. For the cuprates which have a $d$-wave gap symmetry the evidence is that, instead, antiferromagnetic spin fluctuations\cite{carbotte:2011,hwang:2006,hwang:2008c} play the dominant role. The iron pnictides are multiple band systems with an $s$-wave gap which could be anisotropic in a given band\cite{odonovan:1995a,leavens:1971,odonovan:1995} but with an overall sign change between bands\cite{hirschfeld:2011,chubukov:2012} consistent with a spin fluctuation mechanism. While the gap symmetry is different from that in the cuprates, the pnictides exhibit many similarities with the cuprates which further suggest that spin fluctuations\cite{yang:2009a} could play an important role in their superconductivity.

To explore the mechanism that leads to pnictide superconductivity, LiFeAs presents a model system. The compound is superconducting with a relatively high transition temperature of $T_c=$ 17.6 K in the stoichiometric parent state, which is unlike most other high $T_c$ materials which need to be doped to become superconducting. This leads to unavoidable disorders\cite{tapp:2008}. No static magnetic order is observed in this material and its spin fluctuation spectrum is much weaker than that of Co-doped Ba-122\cite{qureshi:2012}. A conclusion of an early angular resolved photoemission (ARPES) study\cite{kordyuk:2011} is that the superconductivity in this particular material may be due to the electron-phonon interaction enhanced by electron-electron interactions and the existence of a van Hove singularity which further increases the density of states. Another ARPES study\cite{borisenko:2012} shows an anisotropic gap and concludes that orbital fluctuations plus a large electron-phonon interaction could account for the superconductivity which is multiband in agreement with results from other probes. Optical spectroscopy reveals a pair of clear dirty limit gaps at $2\Delta_0=$ 3.2 meV and 6.3 meV\cite{min:2013}. Scanning tunneling spectroscopy (STM) shows a homogeneous gap structure also with two gaps at $2\Delta_0 =$ 5.0 and 10.6 meV\cite{chi:2012}.

The boson-exchange model has proved to be remarkably successful in accounting for many of the spectroscopic properties of the high temperature superconductors\cite{carbotte:2011} and there is some evidence that this model may also be applicable to the pnictides\cite{chi:2012,yang:2009,hanaguri:2012}. To test the model, one extracts the bosonic spectrum from experimental data on the inelastic scattering and compares this with known neutron scattering data on the same material. Thus one may be able to identify the bosons as spin fluctuations or phonons by their spectroscopic signatures.

There are many ways to get information about the inelastic scattering including angle-resolved photoemission (ARPES)\cite{garcia:2010,shen:1995,damascelli:2003,zhang:2008,bok:2010,schachinger:2008}, Raman scattering\cite{muschuler:2010}, scanning tunneling spectroscopy (STS)\cite{hanaguri:2012}, and in particular infrared absorption (IR)\cite{carbotte:2011,hwang:2006,hwang:2008c,yang:2009a,yang:2009}. Recently STS tunneling data has appeared for LiFeAs\cite{chi:2012,hanaguri:2012} which has revealed a boson structure in the superconducting density of states at an energy around 7 meV\cite{chi:2012}. Inelastic neutron scattering of LiFeAs also shows an increase in scattering around $E = 8$ meV on cooling below $T_c$\cite{taylor:2012}. Other data are consistent with these observations\cite{qureshi:2012,wang:2012}. It appears that in some pnictides there is a close relationship between the structure seen in the STS data, as an example for Ba$_{0.6}$K$_{0.4}$Fe$_2$As$_2$ and for Na(Fe$_{0.975}$Co$_{0.025}$)As\cite{wang:2012a}, and a magnetic resonance at $\Omega_R^{INS} \cong$ 4.3 $k_B T_c$\cite{wang:2012a,lumsden:2009,inosov:2010}. This is reminiscent of the situation in the high $T_c$ oxides where coupling to a resonance mode is observed in the optics at $\Omega^{OPT}_R \sim$ 6.3 $k_B T_c$ which is close to the canonical value of the spin resonance seen in neutron scattering $\Omega^{INS}_R \sim$ 5.4 $k_B T_c$\cite{he:2001,he:2002}. Specific to LiFeAs is the observation\cite{qureshi:2012} of incommensurate $\vec{q} = \vec{Q}_{inc}$ spin fluctuations which can be described by a form $\Gamma \omega/(\omega^2+\Gamma^2)$ with $\Gamma = 6.0 \pm 0.6$ meV. It should be emphasized that in optics one deals with an average over all momenta and not just with a particular $\vec{Q}_{inc}$. Also, optical data\cite{min:2013} have very recently become available, but up to now the data has not been analyzed in terms of an electron-boson exchange model to recover the electron-boson spectral density $I^2\chi(\omega)$ associated with inelastic scattering. Here we provide such an analysis. In section 2 we present the formalism needed to extract an electron-boson spectral density from the optical data. this involves a generalization of the usual maximum entropy technique\cite{schachinger:2006} to the case of a superconducting $s$-wave gap. In section 3 we present our results. The need to include anisotropy is discussed in section 4 and conclusions are drawn in section 5.

\section{Formalisms}

A first step in such an analysis has often\cite{carbotte:2011}, but not always\cite{heumen:2009,heumen:2009b} been to start from optical scattering rates, $1/\tau^{op}(\omega)$, introduced in optics in direct analogy to the more familiar quasiparticle scattering rates, $1/\tau^{qp}(\omega)$, which describe the broadening of the electron spectral function. One starts with the optical conductivity $\sigma(\omega)$\cite{carbotte:1995} and defines $1/\tau^{op}(\omega) = \frac{\Omega_{pl}^2}{4 \pi} Re\Big{\{} \frac{1}{\sigma(\omega)}\Big{\}}$ where $\Omega_{pl}$ is the plasma frequency. Using ordinary perturbation theory in a boson-exchange model, Allen\cite{allen:1971} has given a simple but remarkable quantitative relationship between the optical scattering rate and the electron-boson spectral density with including a residual optical impurity scattering $1/\tau_{imp}$,
\begin{equation}\label{eq1}
\frac{1}{\tau^{op}(\omega, T)}=\int_{0}^{\infty}\!\!d\Omega\:\: K(\omega,\Omega, T) I^2\chi(T,\Omega) + \frac{1}{\tau_{imp}},
\end{equation}
with the kernel $K(\omega, \Omega, T)$, given in the normal state at finite temperature ($T$) by\cite{shulga:1991}
\begin{eqnarray}\label{eq2}
K(\omega,\Omega, T) &=& \frac{\pi}{\omega}\Big{[} 2\omega \coth\Big{(} \frac{\Omega}{2T} \Big{)} -(\omega+\Omega) \coth\Big{(} \frac{\omega+\Omega}{2T} \Big{)} \nonumber \\  &+& (\omega-\Omega) \coth\Big{(} \frac{\omega-\Omega}{2T} \Big{)} \Big{]},
\end{eqnarray}
and in the superconducting state with isotropic\cite{allen:1971,schachinger:2006} $s$-wave gap (anisotropy neglected\cite{odonovan:1995a,leavens:1971,odonovan:1995b}) in the clean limit by
\begin{eqnarray}\label{eq3}
K(\omega,\Omega, T=0) &=& \frac{2\pi}{\omega} (\omega-\Omega) \theta(\omega-2\Delta_0-\Omega) \nonumber\\
&\times&E\Big{(} \sqrt{1-\frac{4\Delta_0^2}{(\omega-\Omega)^2}} \Big{)}.
\end{eqnarray}
Here $E(x)$ is the complete elliptical integral of the second kind, $\theta(x)$ is the Heavyside function with $\theta(x < 0)$ = 0 and $\theta(x \geq 0)$ = 1, and $\Delta_0$ is the isotropic gap. More complicated formulas have also appeared for the case in which an energy dependent density of states $N(\omega)$ plays an important role\cite{sharapov:2005,mitrovic:1983:2}, but these are not required here. It is sufficient to make the point that at zero temperature in the normal state the modified kernel has the form
\begin{equation}\label{eq4}
\frac{1}{\tau^{op}(\omega, T=0)}=\frac{2 \pi}{\omega}\int_{0}^{\omega}d\Omega I^2\chi(\Omega)\int_{0}^{\omega-\Omega}d\omega'\tilde{N}(\omega'),
\end{equation}
where $\tilde{N}(\omega')$ is the Fermi surface symmetrized band structure electronic density of states $[N(\omega')+N(-\omega')]/2N(0)$. For the specific case of LiFeAs the electronic band structure calculations of $\tilde{N}(\omega)$ from the data given by Nekrasov {\it et al.}\cite{nekrson:2008} show that as long as our inversions are limited to low energies $\leq$ 40 meV, to a good approximation we can take $\tilde{N}(\omega) = 1.0$, a constant.
To include impurities a residual optical impurity scattering $1/\tau_{imp}$ was added to the right hand side of Eq. (\ref{eq1}) for the normal state and this same value times the elliptic integral $E( \sqrt{1-4\Delta_0^2/(\omega-\Omega)^2})$ of Eq. (\ref{eq3}) evaluated at $\Omega = 0$ for the superconducting state\cite{allen:1971}.

In the perturbation theory approach of Allen\cite{allen:1971} the charge carrier-boson spectral density $I^2\chi(\Omega)$ that enters Eq. (\ref{eq1}) is naturally its transport version. In transport there is an extra vertex factor which selectively weight more heavily backward as compared to forward scattering events which are eliminated as they do not deplete the current. The same formula, Eq. (\ref{eq1}), can also be obtained from an Eliashberg formulation with neglect of vertex corrections as was shown by Shulga {\it et al.}\cite{shulga:1991} and others. It is such vertex corrections addressed in the seminal work of Gotze and Wolfle\cite{gotze:1972}, which have the effect of changing the spectral function $I^2\chi(\Omega)$ from its quasiparticle to its transport form. For a more modern discussion of vertex corrections, refer to the work of Cappelluti and Benfatto\cite{cappelluti:2009} who treat the specific case of graphene. Any further small corrections which cannot be incorporated within the definition of an effective electron-boson transport spectral density are not considered here as they would be specific to a particular microscopic model, and this is not suitable for the analyses of data as we wish to do in this paper.

A detailed analysis of the accuracy and limitations of formulas (\ref{eq1}) to (\ref{eq3}) is found in Ref. \cite{schachinger:2006} where a maximum entropy technique is also described which allows one to recover the electron-boson spectral function $I^2\chi(\Omega)$ from a knowledge of the scattering rates $1/\tau^{op}(\omega)$. A summary of this technique is given as follows. Eq. (\ref{eq1}) can be discretized $D_{in}(i) = \sum_j K(i,j) I^2\chi(j) \Delta\Omega$ with $\Omega_j = j\Delta\Omega$, $\Delta \Omega$ the grid size on $\Omega$, and $j$ an integer. We define a $\chi^2$ by
\begin{equation}\label{eqa1}
\chi^2 = \sum^N_{i = 1}\frac{[D_{in}(i)-1/\tau^{op}(i)]^2}{\sigma_i^2}
\end{equation}
where $D_{in}(i)$ is the input data for the optical scattering rate. $1/\tau^{op}(i)$ is defined by Eq. (\ref{eq1}) and is a functional of $I^2\chi(\omega)$ and $N$ is the number of data points available. Here $\sigma_i$ is the error assigned to the data $D_{in}(i)$. The entropy functional
\begin{equation}\label{eqa2}
L = \frac{\chi^2}{2} - a S
\end{equation}
is minimized with the Shannon-Jaynes entropy, $S$\cite{schachinger:2006}
\begin{equation}\label{eqa3}
S = \int_0^{\infty}d\Omega\Big{[} I^2\chi(\Omega)-m(\Omega)-I^2\chi(\Omega)\ln\Big{|}\frac{I^2\chi(\Omega)}{m(\Omega)} \Big{|} \Big{]}.
\end{equation}
which gets maximized in the process. The determinative parameter $a$ in Eq. (\ref{eqa2}) controls how close the fit should follow the data while not violating the physical constraints on $I^2\chi(\omega)$. Furthermore, $m(\Omega)$ is the constraint function (default model) which is taken to be some constant value indicating that there is no a priori knowledge of the functional form of the electron-boson density $I^2\chi(\Omega)$. Finally, we will make use of the historical maximum entropy method which iterates $a$ until the average $<\chi^2> = N^2$ is achieved with acceptable accuracy.

\section{Results and discussions}

\begin{figure}[t]
  \vspace*{-0.5 cm}%
  \centerline{\includegraphics[width=4.5 in]{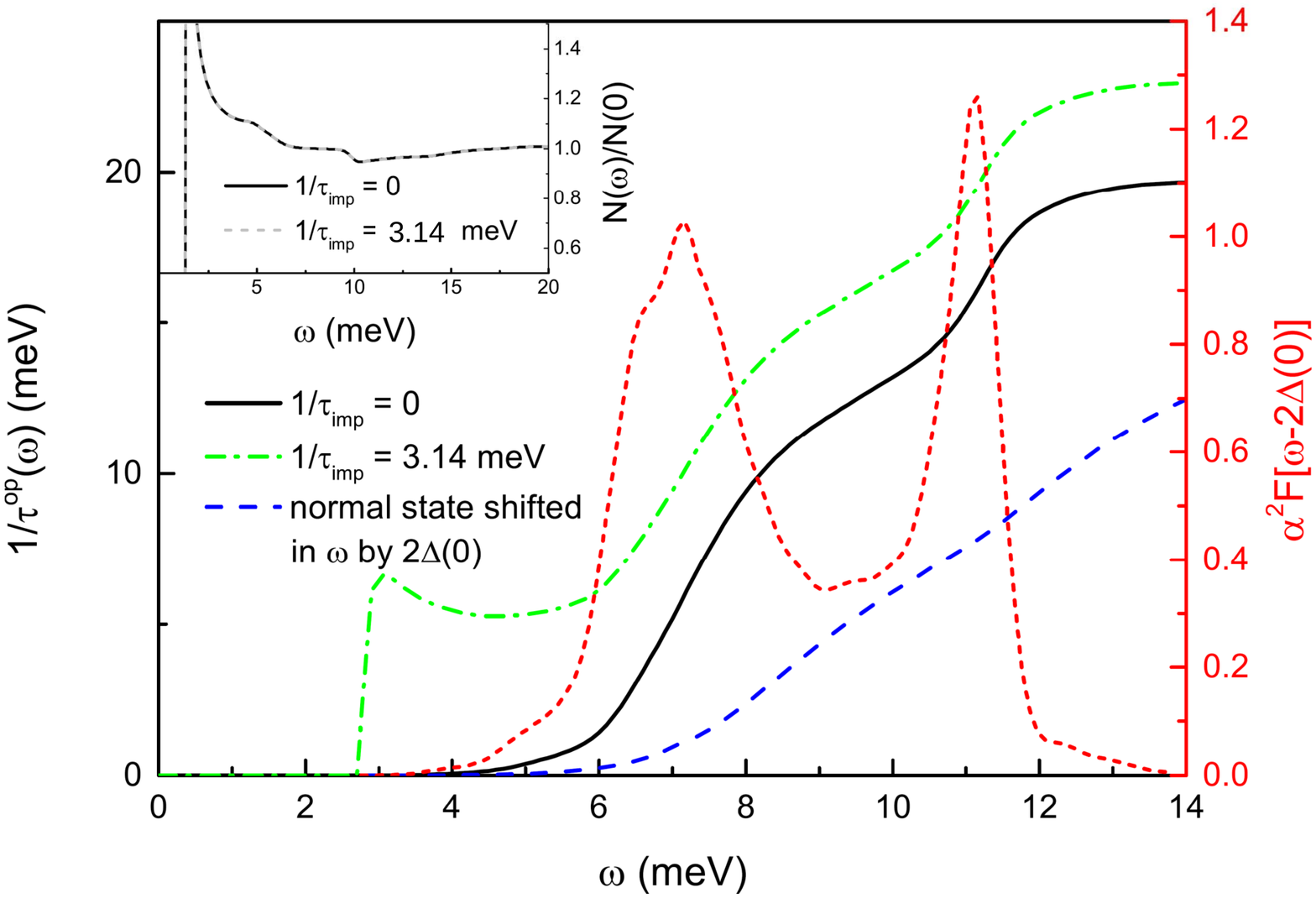}}%
  \vspace*{-0.5 cm}%
\caption{(Color online) The optical scattering rate $1/\tau^{op}(\omega)$ in meV as a function of energy $\omega$ (in meV) calculated in isotropic Eliashberg theory based on the Pb electron-phonon spectral density $\alpha^2F(\omega)$ shown in short dashed red curve (right hand sale applies). The dashed blue curve is the normal state without impurity scattering displaced along the horizontal axis by twice the gap. It is to be compared with the solid black curve which is the superconducting case (pure limit). Including impurity scattering with $1/\tau_{imp} =$ 3.14 meV gives the dash-dotted green curve. The inset shows the normalized superconducting density of states $N_s(\omega)$ with (dashed grey) and without (solid black) impurity scattering. The curves are identical showing that elastic scattering has no effect on the superconducting density of states.}
 \label{fig1}
\end{figure}

Fig. \ref{fig1} sets the stage for what is to come next. It is based on the electron-phonon spectrum $\alpha^2F(\omega)$ [dotted red spectrum] of Pb obtained by an inversion of tunneling data based on the full Eliashberg\cite{carbotte:1990,mcmillan:1965} equations. In the inset we show the electronic density of states $N(\omega)/N(0)$ for the superconducting state at zero temperature ($T =$ 0) where the boson structure due to the two prominent peaks in $\alpha^2F(\omega)$ is clearly seen. It is also seen that these structures are independent of residual impurity scattering {\it i.e.} $1/\tau_{imp}$ drops out of $N(\omega)$ for an isotropic $s$-wave gap. This is no longer the case when optical characteristics are considered instead as summarized in the main frame where we plot $1/\tau^{op}(\omega)$\cite{carbotte:1995} vs $\omega$ for three cases. The dashed (blue) curve is for pure Pb in the normal state with no residual scattering, and the curve has been displaced along the horizontal axis by twice the gap edge $\Delta_0$ for a more convenient comparison with the solid black curve which gives the result in the superconducting state close to $T =$ 0. Note that the boson structure is more pronounced in this curve than it is in the normal state [dashed blue curve] and consequently, inversion of optical data in this case is more favorable. On the other hand, impurities do strongly affect $1/\tau^{op}(\omega)$ as we can see in the dash-dotted (green) curve in sharp contrast to the density of states for which they drop out. It is important to emphasize, however, that the inclusion of a residual scattering rate of $1/\tau_{imp} =$ 3.14 meV has mainly affected the region immediately above the gap edge at $\omega = 2\Delta(0)\equiv 2\Delta_0 \cong$ 2.78 meV while leaving the phonon structure at higher energies almost unaffected. It is this region that we are mainly interested in.

\begin{figure}[t]
  \vspace*{-0.50 cm}%
  \centerline{\includegraphics[width=4.0 in]{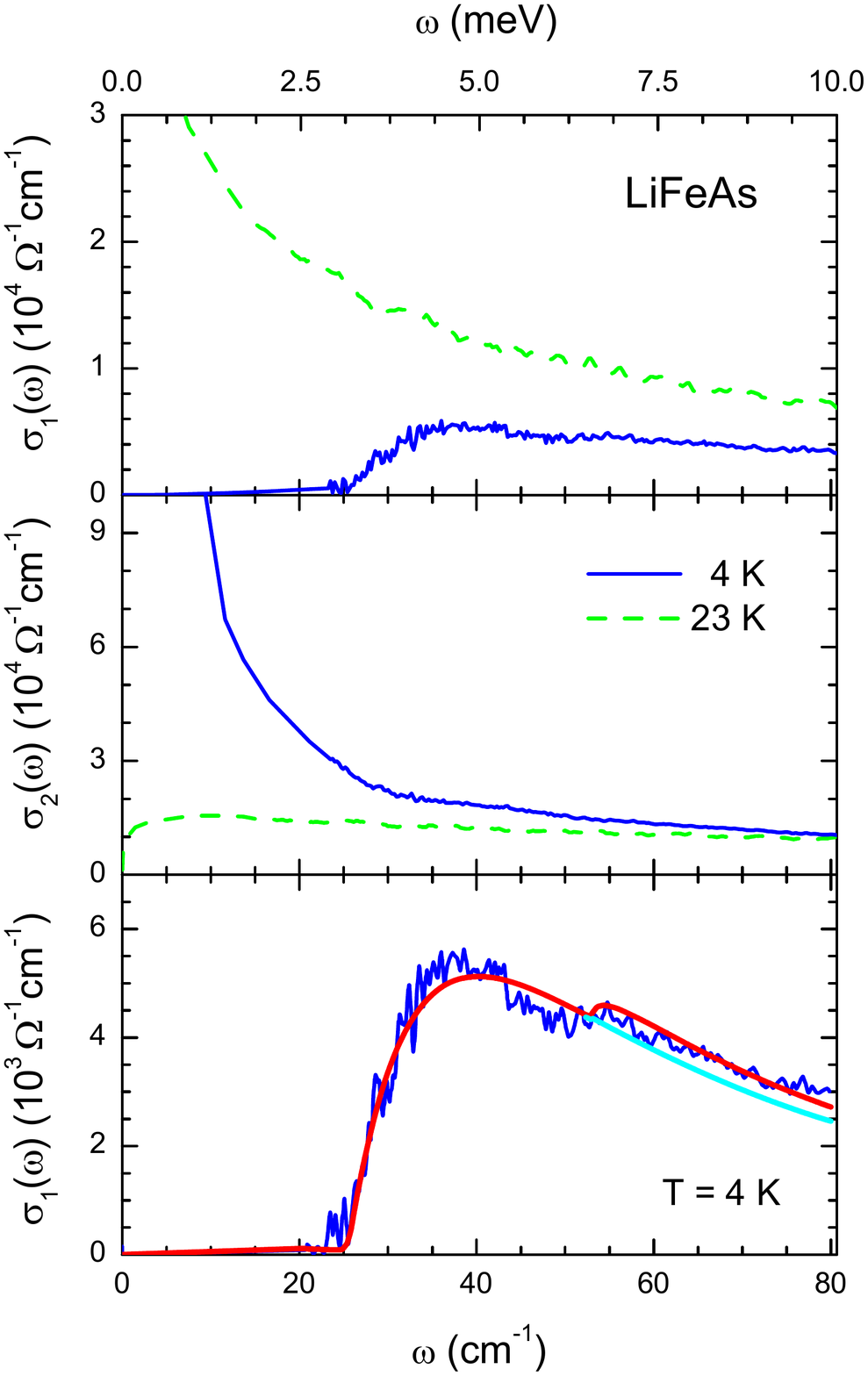}}%
  \vspace*{-0.5 cm}%
\caption{(Color online) Top frame gives the real part of the dynamic conductivity $\sigma_1(\omega)$ in units of 10$^4$ $\Omega^{-1}$cm$^{-1}$ as a function of photon energy $\omega$ in cm$^{-1}$. The green dashed curve is for the normal state at temperature $T =$ 23 K just above $T_c$ and the blue continuous curve applies at $T =$ 4 K in the superconducting state. The middle frame gives the corresponding imaginary parts $\sigma_2(\omega)$. The bottom frame shows the superconducting state data $\sigma_1(\omega)$ (continuous blue curve) and the fit of reference \cite{min:2013} (solid red curve) to a two-independent band generalized Mattis-Bardeen model\cite{zimmermann:1991}.}
 \label{fig2}
\end{figure}

\begin{figure}[t]
  \vspace*{-1.0 cm}%
  \centerline{\includegraphics[width=4.0 in]{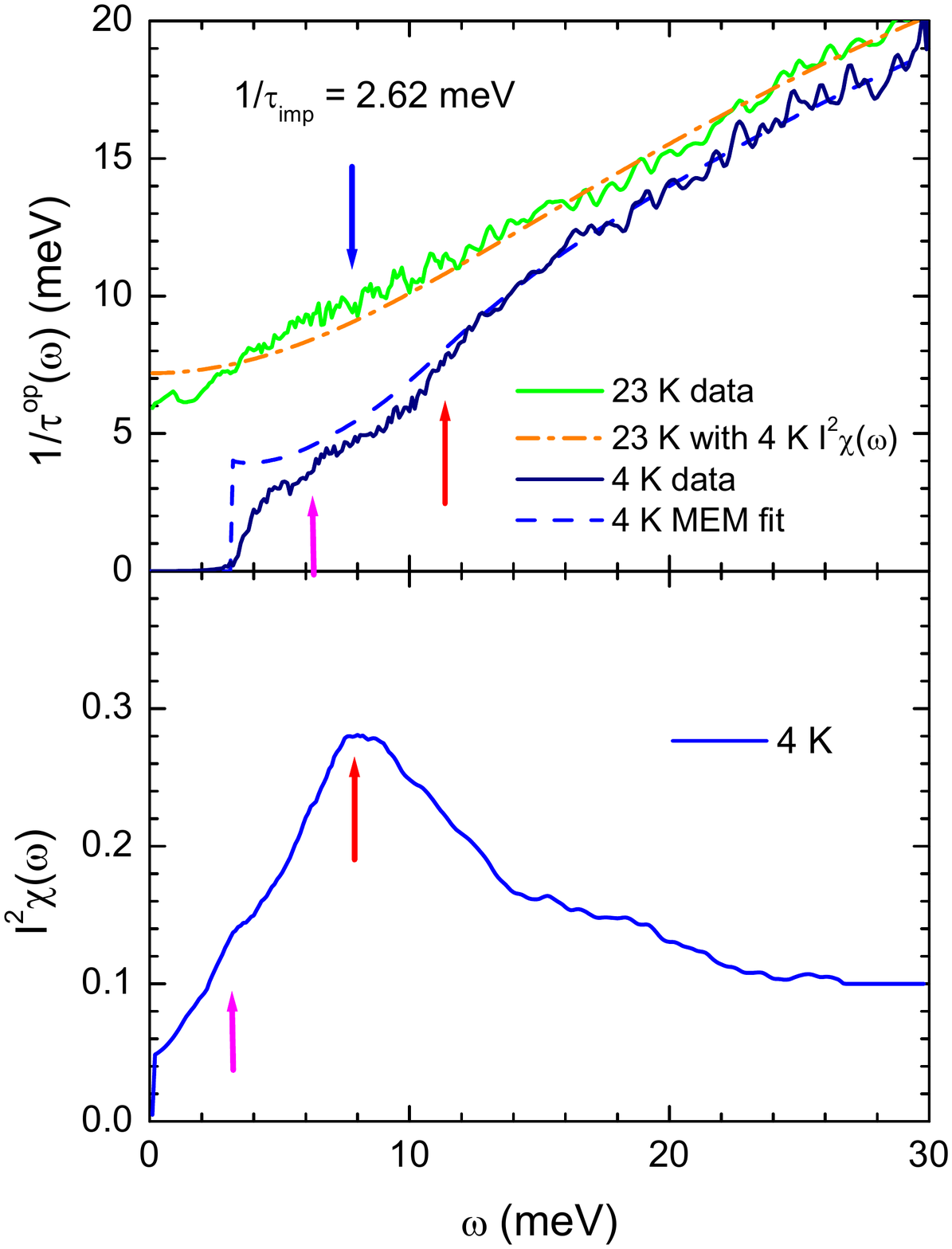}}%
  \vspace*{-1.5 cm}%
\caption{(Color online) The solid curves in the top frame give our experimental results for the optical scattering rate $1/\tau^{op}(\omega)$ in meV as a function of photon energy $\omega$ in meV. The green curve is the normal state data at temperature $T =$ 23 K and the dark blue is the superconducting state data at $T =$ 4 K. The dash-dotted orange curve is our theoretical results in the normal state obtained from the electron-boson spectral density $I^2\chi(\omega)$ of the lower frame which we obtained from a maximum entropy fit (blue dashed curve) to the scattering rate data at $T =$ 4 K with a superconducting isotropic gap value of 1.59 meV and an impurity scattering rate of 2.62 meV. The purple and red arrows emphasize a structure in $I^2\chi(\omega)$ at $\sim$ 3.5 meV and a peak at $\sim$ 8 meV, respectively. These same structures are also highlighted in the upper frame with arrows. Note that the purple and red arrows have been displaced to higher energies by 3.18 meV i.e. $2\Delta_0$.}
 \label{fig3}
\end{figure}

In Fig. \ref{fig2} we show data for LiFeAs from reference \cite{min:2013} on the real part (top frame) and imaginary part (middle frame) of the dynamic conductivity $\sigma_1(\omega)$ and $\sigma_2(\omega)$ in units of 10$^4$ $\Omega^{-1}$cm$^{-1}$ as a function of photon energy $\omega$ for two temperatures. The green dashed curve is the data in the normal state at temperature $T =$ 23 K, while the blue continuous curve is in the superconducting state at $T =$ 4 K with the critical temperature $T_c =$ 17.6 K. The bottom frame shows a fit (solid red curve) to the data for the superconducting case (solid blue curve) for $\sigma_1(\omega)$ in units of 10$^{3}$ $\Omega^{-1}$cm$^{-1}$. The fit is based on the generalized Mattis-Bardeen formula\cite{zimmermann:1991} for the conductivity in a BCS approach. It provides a two-independent band picture with a gap of $\Delta_0 =$ 1.59 meV and residual scattering rate $1/\tau_{imp} =$ 4 meV and a larger gap $\Delta_0 =$ 3.3 meV and scattering rate $1/\tau_{imp} =$ 1 meV corresponding to the cleaner band. This second band makes only a very minor contribution to the total conductivity up to 10 meV as indicated by the difference between red curve and solid light blue curve above $\sim$ 6 meV. As we will be interested in inverting only the low energy data a single band picture should apply and we expect the recovered electron-boson spectral density that we get from a maximum entropy fit, to be characteristic of the band with the smaller gap. Results of the inversion for $I^2\chi(\omega)$ are shown in the lower frame of Fig. \ref{fig3} as the solid blue curve up to 30 meV. We see a shoulder structure around 3.5 meV as well as a prominent peak at $\sim$ 8 meV. We cannot rule out that the region above this energy has a small contribution from the second band and hence that our recovered spectral density at these higher energies is representative of an average over the two bands. Here however our main interest is the low energy part of $I^2\chi(\omega)$. In the fit shown the residual scattering was $1/\tau_{imp} =$ 2.62 meV which makes a contribution of 2.62$\times E(\sqrt{1-4\Delta_0^2/\omega^2})$ to the total (residual plus inelastic) scattering rate in the superconducting state at $T =$ 4 K. We have assumed a constant gap value of $\Delta_0 =$ 1.59 meV and $T =$ 0 in our fitting procedure. The upper frame of Fig. \ref{fig3} shows the data for the scattering rate at $T =$ 4 K (solid dark blue curve) and our maximum entropy fit (dashed blue curve). Except for the region near the gap edge at $2\Delta_0 =$ 3.18 meV the fit is good. The discrepancy in the region immediately above $\omega = 2\Delta_0$ is attributed to our neglect of gap anisotropy as we will discuss later. The purple and red arrows serve to emphasize the shoulder structure and the position of the peak respectively. In the scattering rate these structures are shifted to higher energies by $2\Delta_0$ of course. The second set of curves in the upper frame for the optical scattering rate apply to the normal state at $T =$ 23 K. The solid green curve is the data obtained from $\sigma_1(\omega)$ and $\sigma_2(\omega)$ while the dash-dotted orange curve is theory. It is obtained using formula Eq. (\ref{eq1}) with the finite temperature kernel Eq. (\ref{eq2}) and the same spectrum $I^2\chi(\omega)$ in the superconducting state shown in the lower frame. We see that the fit is reasonable and we emphasize that there is no need to modify the electron-boson spectral density with increasing temperature. To calculate scattering rates we needed an estimate of the plasma energy $\Omega_{pl}$ for the band of interest in addition to $\sigma_1(\omega)$ and $\sigma_2(\omega)$: $1/\tau^{op}(\omega, T) = \Omega_{pl}^2 \: \sigma_1(\omega, T)/ \{4\pi\:[\sigma_1(\omega, T)^2+\sigma_2(\omega, T)^2]\}$. In a multiband system it is difficult to determine from optical data alone so we adjusted $\Omega_{pl}$ to 1.25 eV to get from our $I^2\chi(\omega)$ spectrum the measured gap of 1.59 meV using Eliashberg theory.

\begin{figure}[t]
  \vspace*{-0.5 cm}%
  \centerline{\includegraphics[width=4.5 in]{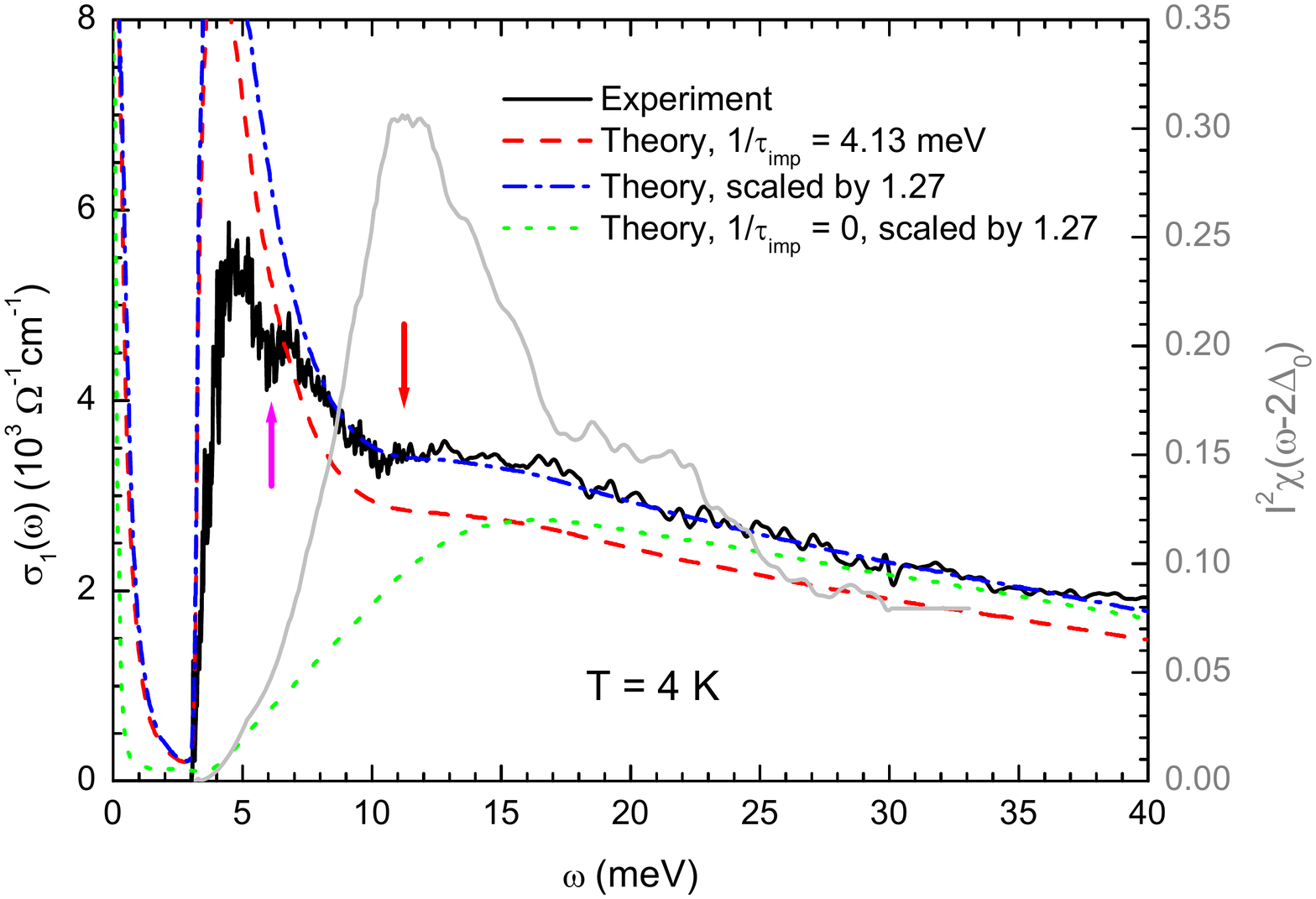}}%
  \vspace*{-0.5 cm}%
\caption{(Color online) The real part of the optical conductivity $\sigma_1(\omega)$ in 10$^3$ $\Omega^{-1}$cm$^{-1}$ as a function of photon energy $\omega$ in meV up to 40 meV. The heavy black curve is the data of Ref. \cite{min:2013}. The dash-dotted blue line is the theoretical result that we obtained by multiplying a factor 1.27 to the dashed red curve which we got using the recovered $I^2\chi(\omega)$ shown as the grey curve (right hand scale applies) in a full Eliashberg calculation of the conductivity for an impurity scattering rate $1/\tau_{imp} =$ 4.13 meV. The dotted green curve is for comparison and is the result for $\sigma_1(\omega)$ when no impurity scattering is included in the numerical work. The heavy red arrow emphasizes the onset of Holstein incoherent processes resulting from electron-boson scattering. The purple arrow emphasizes the structure at $\sim$ 6.5 meV captured by the shoulder structure at $\sim$ 3.3 meV (6.5 - 2$\Delta_0$) in $I^2\chi(\omega)$ of the lower frame of Fig. \ref{fig3} but not included in the simplified model $I^2\chi(\omega)$ used here.}
 \label{fig4}
\end{figure}

The peak in $I^2\chi(\omega)$ that we have obtained is also seen directly in the real part of the optical conductivity $\sigma_1(\omega)$. This is demonstrated in Fig. \ref{fig4}. The heavy solid black curve gives $\sigma_1(\omega)$ in $10^3$ $\Omega^{-1}$cm$^{-1}$ from Ref. \cite{min:2013}. We emphasize with a heavy red arrow the dip in this data just above 10 meV. We take this as the onset of the Holstein incoherent boson assisted processes which set in at an energy of $2\Delta_0 + \Omega_R$ where $\Omega_R$ is the peak position in $I^2\chi(\omega)$ (solid grey line, right hand scale applies). The other curves shown are results of full Eliashberg calculations of the real part of $\sigma(\omega)$ in the superconducting state based on this model electron-boson spectrum (solid grey line). it is only slightly different from the$I^2\chi(\omega)$ shown in the lower frame of Fig. \ref{fig3}. It was obtained in a somewhat different fit to the optical data with a slightly larger value of residual scattering namely $1/\tau_{imp} =$ 4.13 meV. Note that the model spectra for $I^2\chi(\omega)$ of Fig. \ref{fig4} and Fig. \ref{fig3} (lower frame) differ only below 4 meV where the spectrum in Fig. \ref{fig3} has a shoulder and more spectral weight than does that in Fig. \ref{fig4}. This serves to illustrate that the position, shape and amplitude of the main peak at $\sim$ 8 meV is robust and not dependent on the details of our fitting procedure. Further, an increase in $1/\tau_{imp}$ can be largely compensated for by an increase in spectral weight in the resulting $I^2\chi(\omega)$ at small $\omega$ with no overall qualitative change to the fit. The dotted green curve is our result for $\sigma_1(\omega)$ in the pure case and the dash-dotted blue curve includes a residual scattering of $1/\tau_{imp} =$ 4.13 meV. Both have been scaled up by a factor of 1.27 for easier comparison with the data. Without this factor we got the dashed red curve which falls somewhat below the data but clearly shows the same Holstein onset as does the data. As a final comment the second purple arrow in Fig. \ref{fig4} shows a structure at $\sim$ 6.5 meV which reflects the shoulder structure shown in the lower frame of Fig. \ref{fig3} for $I^2\chi(\omega)$ not included in the calculation of $\sigma_1(\omega)$ shown in Fig. \ref{fig4}. This shoulder could be an artifact associated with the onset of the second gap (see lower frame of Fig. \ref{fig2}).

\section{Inclusion of anisotropic gap}

\begin{figure}[t]
  \vspace*{-0.5 cm}%
  \centerline{\includegraphics[width=4.0 in]{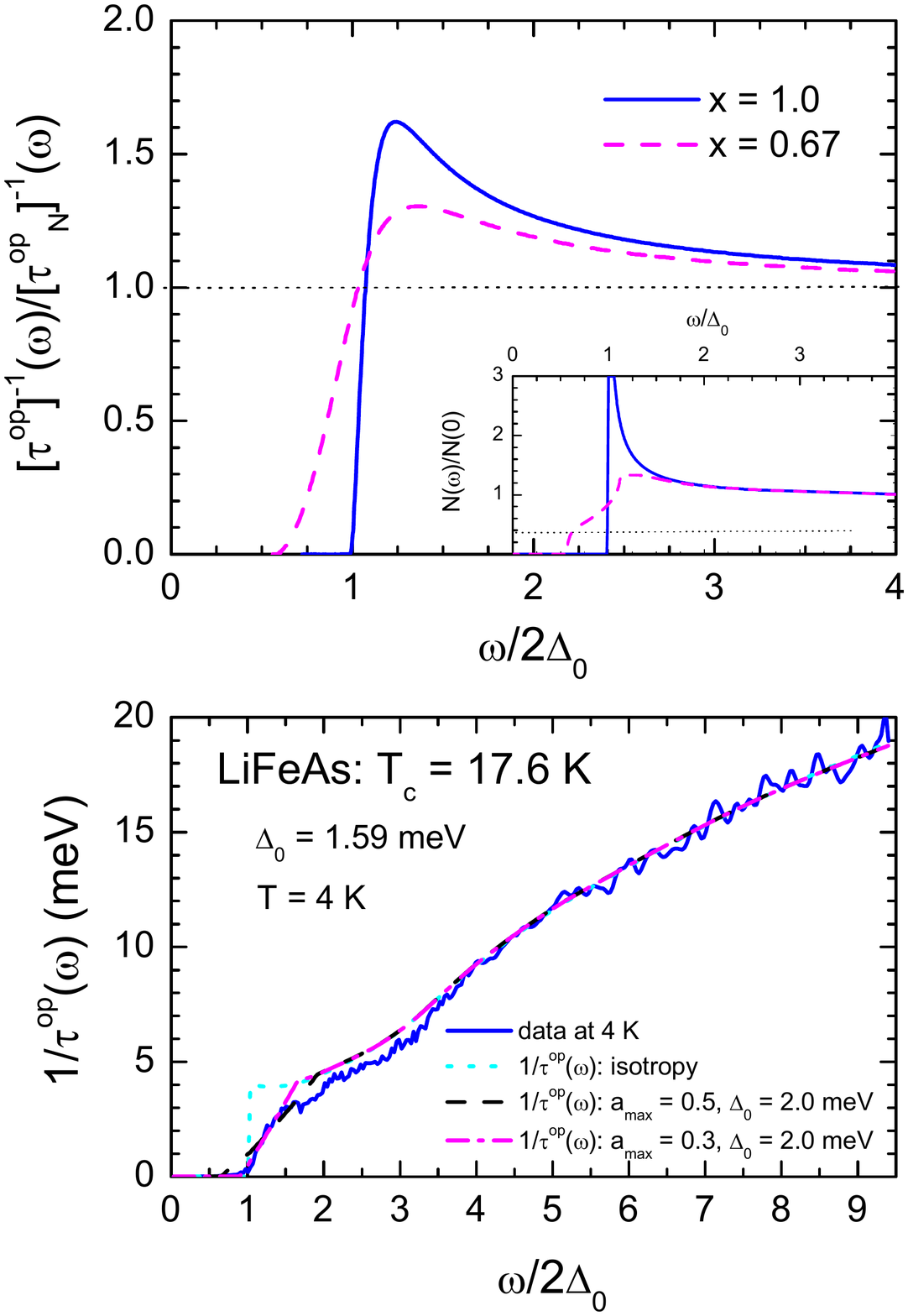}}%
  \vspace*{-1.0 cm}%
\caption{(Color online) In top frame we display the optical scattering rate normalized to its normal state value obtained in a BCS model with $x =$ 0.67 (dashed purple line) and $x =$ 1.0 (solid blue line) without gap anisotropy in a $s$+$d$ wave model as specified in text [Eq. (\ref{eq5}) and Eq. (\ref{eq5a})]. The inset shows the corresponding superconducting state density of states. In bottom frame we display the optical scattering rate $1/\tau^{op}(\omega)$ in meV as function of $\omega/2\Delta_0$ compared with data (solid blue curve). The short dashed light-blue fit is for an isotropic gap $\Delta_0 =$ 1.59 meV. The dashed black includes anisotropy with $a_{max} =$ 0.5 and gap $\Delta_0 =$ 2.0 meV while the dash-dotted purple curve is for $a_{max} =$ 0.3 in a model where the gap is distributed with equal probability between $\Delta_0(1-a_{max})$ and $\Delta_0(1+a_{max})$.}
 \label{fig5}
\end{figure}

Next we turn to the question of gap anisotropy. As we have already noted, the fit in the top frame of Fig. \ref{fig3} in the region just above the gap edge is not good. This points to the importance of gap anisotropy. This anisotropy has been noted by Borisenko {\it et al.}\cite{borisenko:2012} who found significant variation in the value of the gap ($\Delta$) as a function of angle $\phi$ about the large hole two-dimensional Fermi surface at the $\Gamma$ point. A variation of the form $\Delta \sim \Delta_0+\Delta_1 \cos(4\phi)$ is suggested from their angular resolved photoemission data at low temperature ($\sim$ 1 K). Anisotropy in the gap of order of 30 \% was also found in the work of Umezawa {\it et al.}\cite{umezawa:2012}. The effect of gap anisotropy in the optical scattering rate of a BCS superconductor with an anisotropic gap has been studied before by Carbotte and Schachinger\cite{schachinger:2010a} and also features in the work of Wu {\it et al.}\cite{wu:2010a}. They use a model for the gap which is a mixture of an $s$-wave component and a $d$-wave component of the form
\begin{equation}\label{eq5}
\Delta = \Delta_0[\alpha + \sqrt{1-\alpha^2}\sqrt{2}\cos(2\phi)].
\end{equation}
The limit $\alpha =$ 1 is pure $s$-wave while $\alpha =$ 0 is pure $d$-wave. A second parameter $x$ is often used instead of $\alpha$ and defined as
\begin{equation}\label{eq5a}
x = \alpha/[\alpha+\sqrt{1-\alpha^2}]
\end{equation}
which is a direct measure of the relative amount of $s$ character in the gap function. In the top frame Fig. \ref{fig5} we present results, based on reference \cite{schachinger:2010a}, for the optical scattering rate normalized to its normal state value as function of normalized photon energy $\omega/2\Delta_0$ in the case when $x =$ 0.67 as the dashed red curve which is to be compared with the solid blue curve which applies in the isotropic case. In the inset the usual BCS square root singularity at $\omega = 2\Delta_0$ is clearly seen in the isotropic limit (solid blue curve) and this is very significantly smeared out in the anisotropic case (dashed purple curve), and there is significant density of states below the gap at $\omega = 2\Delta_0$. This has the effect of reducing the peak in the optical scattering rate at the gap. Of course some of the details of the chosen model for gap anisotropy could have some effect on the onset of the scattering in the region of the gap but on the whole these details get averaged out when the angular average over $\phi$ is taken to get $1/\tau^{op}(\omega)$. For example the residual scattering rate is given by
\begin{equation}\label{eq6}
\frac{1}{\tau_{imp}(\omega)} = \frac{1}{\tau_{imp}} \int_{1-a_{max}}^{1+a_{max}} E\Big{(}\sqrt{1-\frac{4\Delta_0^2 a^2}{\omega^2}}\Big{)} P(a) da
\end{equation}
in a model for gap anisotropy where $P(a)$ gives the probability that the gap is $\Delta_0(1+a)$. Results for a constant probability $P(a) = 1/2a_{max}$ are presented in the lower frame of Fig. \ref{fig5}. The solid blue curve is the measured scattering rate in units of meV and the light blue short-dashed curve is for the isotropic gap case and is for comparison. For the dash-dotted red curve $a_{max}$ was taken to be 0.3 and gap was readjusted upward to a value of $\Delta_0 =$ 2.0 meV. The dashed black curve is the same but $a_{max}$ has been increased to 0.5. It is clear that including anisotropy has greatly improved our fit to the optical scattering rate data but the details of the anisotropy do not matter much because optics deals with an average over momentum. Of course ARPES can give much more detailed information on its $\phi$ dependence around the Fermi surface. Nevertheless it is clear from our work that LiFeAs has an anisotropic gap.

\begin{figure}[t]
  \vspace*{-0.0 cm}%
  \centerline{\includegraphics[width=4.0 in]{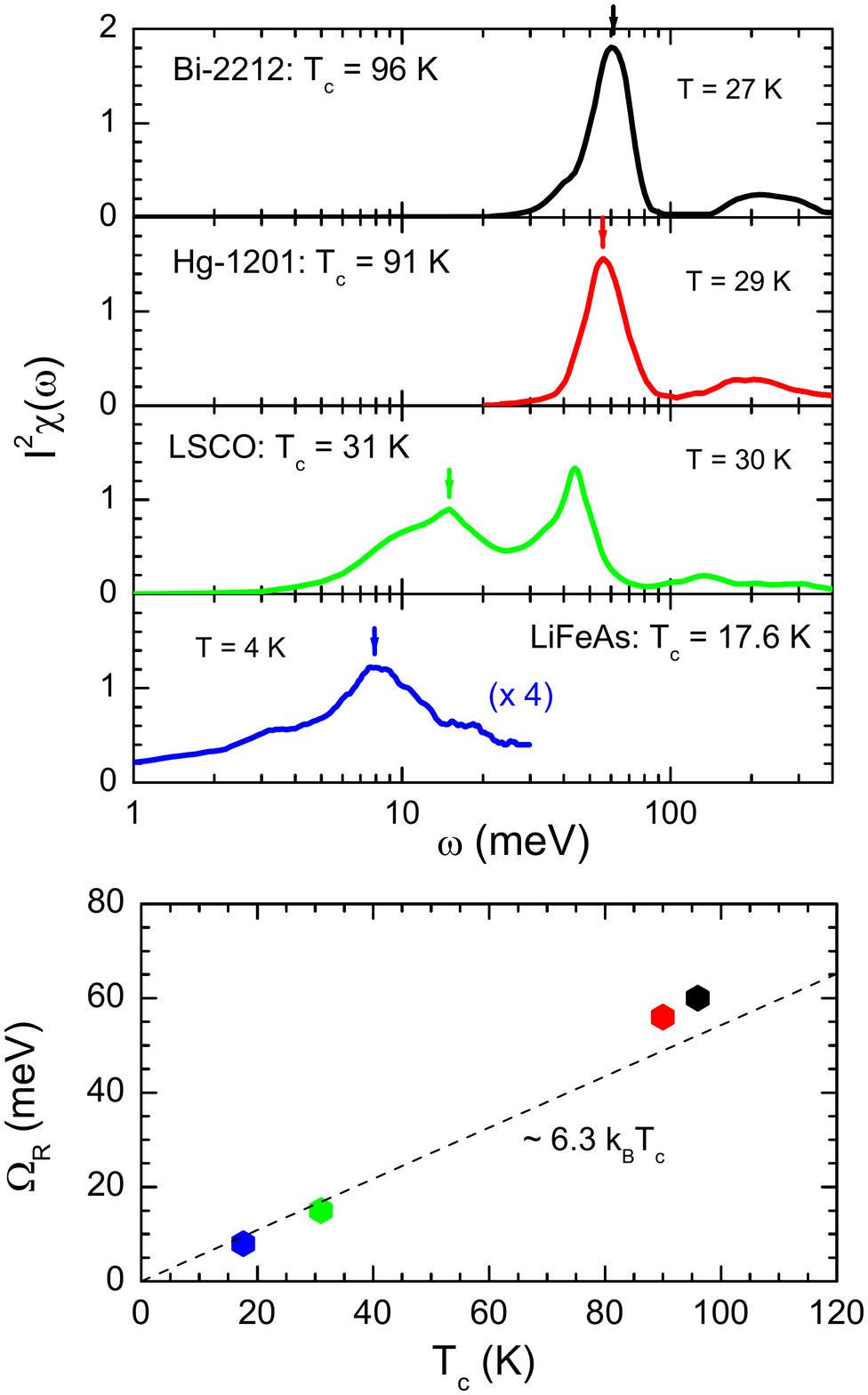}}%
  \vspace*{-0.5 cm}%
\caption{(Color online) In the top panel we display electron-boson spectral density $I^2\chi(\omega)$ obtained from optics in three cuprates, Bi-2212 (black solid curve) with $T_c =$ 96 K in the top frame, Hg-1201 (red solid curve) with $T_c =$ 91 K in the second frame and LSCO (green solid curve) with $T_c =$ 31 K in the third frame, compared in the forth frame with our results for LiFeAs (blue solid curve) with $T_c =$ 17.6 K. The bottom panel shows the value of the lower peak in $I^2\chi(\omega)$ (from optics) in meV as a function of the critical temperature $T_c$ in Kelvins. The dashed line gives $\Omega_R^{OPT} \cong$ 6.3 $k_B T_c$.}
 \label{fig6}
\end{figure}

In Fig. \ref{fig6} we present a comparison of the electron-boson spectral density $I^2\chi(\omega)$ obtained in LiFeAs with equivalent results for the high $T_c$ cuprates. The top frame gives the spectral density obtained in reference \cite{hwang:2007} for Bi-2212 with a $T_c =$ 96 k at $T =$ 27 K in the superconducting state (solid black curve). The next frame is for Hg-1201\cite{yang:2009} with $T_c =$ 91 K (red solid curve). The two spectra are very similar with a sharp peak at $\Omega_R$ marked by an arrow around 60 meV followed at larger energies with a second broader and weaker contribution extending to high energies of order 400 meV and even higher energies up to 2.2 eV\cite{hwang:2014} (not shown here). The third frame is for LSCO with $T_c =$ 31 K\cite{hwang:2008c} and is shown to illustrate that not all spectra characteristic of the high $T_c$ oxides have the same shape. Here there are two peaks and energy of the lower peak is much reduced as compared to the upper two frames. This spectrum is not so different from the $I^2\chi(\omega)$ obtained in this work for LiFeAs shown in the fourth frame of the top panel. However, note in particular that the scale on vertical axis is considerably smaller for LiFeAs than it is for LSCO. There is a factor of roughly 4. The quantity itself is dimensionless. Also, note that the lower frame has considerably more spectral weight below the peak indicated by the arrow than does the third frame for LSCO. In the bottom panel we display that the optical resonance peak energy $\Omega_R$ increases with the value of $T_c$. The four points shown follow $T_c \cong$ 6.3 $k_B T_c$ which applies well to many more oxides complied in Fig. 4 of reference \cite{yang:2009}. This observation, however, does not by itself prove a common origin for the excitations involved in the inelastic scattering reflected in the derived electron-boson spectral function. There is considerable evidence that in the cuprates it is mainly spin fluctuations with perhaps a 10 to 20 \% contribution from phonons as reviewed by Carbotte, Timusk and Hwang\cite{carbotte:2011}. Further the neutron resonance associated with spin fluctuations follows a closely related behavior $\Omega^{INS}_R \cong$ 5.4 $K_B T_c$\cite{he:2001,he:2002}. While it may seem reasonable to assume that LiFeAs simply follows the same trend this may not be the case. Recall that this material shows no static magnetic structure and that the intensity of the observed spin fluctuation spectrum at the incommensurate wave vector $\vec{Q}_{inc} =$ (0.5 $\pm \delta$, 0.5 $\pm \delta$) with $\delta \sim 0.07$ is much reduced\cite{qureshi:2012} by an order of magnitude as compared with that observed in the Co-doped Ba-122 system\cite{lumsden:2009,inosov:2010}. Further the lack of nesting between hole at the $\Gamma$ point and electron pockets at the $M$ point is not favorable to magnetic interaction. Nevertheless Wang {\it et al.}\cite{wang:2013} were able to show, through a detail analysis of the pairing vertex, that spin fluctuation pairing can still be the operative mechanism in this material. We also find no evidence that our peak in $I^2\chi(\omega)$ at $\sim$ 8 meV is significantly depleted as the normal state is approach with increasing temperature. The neutron resonance emphasized in the cuprates is due to superconductivity and is found to vanish at $T_c$ as is also the case in BaFe$_{1.85}$Co$_{0.15}$As$_2$\cite{inosov:2010}. It needs to be kept in mind however that such a spin fluctuation mode which vanishes at $T_c$ does not, on its own, cause the superconductivity. Rather it can reenforce it at lower temperature. Admittedly its observation in an electron spectrum shows that charge carriers are importantly coupled to the spin degree of freedom but the inverse is not necessarily true. As recognized early\cite{carbotte:1999} the spin fluctuations provide as well broad background extending to very high energies and it is the coupling of the charge carriers to this background which makes the major contribution to the pairing glue. It is possible that major part of our recovered spectrum $I^{2}\chi(\omega)$ is the spin fluctuation background.   Based on ARPES, Kordyuk {\it at al.}\cite{kordyuk:2011} have determined an electron-boson spectral density associated with two different cuts on the hole-like two-dimensional Fermi surface around the $\Gamma$ point. They find peaks at $\sim$ 15, 30 and 44 meV which they associate with coupling to phonons and argue that the mechanism which drives the superconductivity may be phonons enhanced by both a van Hove singularity in the density of states near the Fermi energy and by electron-electron interactions. It should be kept in mind, however, that ARPES is momentum specific and that superconductivity depends rather on an average of the electron-boson spectral density for all electrons as is probed directly in optics. Nevertheless it is interesting to note that in the right hand frame of their Fig. 2 the derived spectrum has considerable weight at low energy $\omega$ as we have found in our optically derived spectrum and this is different from the cuprate spectra shown in the top three frames of Fig. \ref{fig6} and could be an indication that a different mechanism is involved. Note that Borisenko {\it et al.}\cite{borisenko:2012} suggest that orbital fluctuations assisted by phonons may play an important role in LiFeAs. What is clear from the lowest frame of the top panel of Fig. \ref{fig6} is that optics shows no clearly recognizable peaks at phonon energies identified in reference\cite{kordyuk:2011}.

\section{Conclusions}

We presented a generalized maximum entropy technique and apply it to the iron based superconductor LiFeAs. This allows us to recover the electron-boson spectral density $I^2\chi(\omega)$ from superconducting state optical data when the gap has $s$-wave symmetry. Anisotropy could also be present. This contrasts with the well-studied case of cuprate high $T_c$ superconductors for which the gap has $d$-wave symmetry and the clean limit applies to first approximation. When the gap instead has $s$-wave symmetry careful consideration of the residual elastic impurity scattering is required because it has a major effect on the onset of the optical absorption at twice the gap value ($2\Delta_0$) and consequently on the optical scattering rate in this energy region. In fact, for an isotropic $s$-wave gap, the optical scattering rate $1/\tau^{op}(\omega)$ has a sharp vertical rise exactly at $\omega = 2\Delta_0$ and this rise is proportional to the impurity scattering rate $1/\tau_{imp}$. The sharp rise in $1/\tau^{op}(\omega)$ at $2\Delta_0$ corresponds to a maximum and is followed by a gradual decrease on the scale of the gap energy before it eventually starts to increase again as the inelastic scattering sets in with modulations in this scattering reflecting structures in the underlying electron-boson spectral density $I^2\chi(\omega)$. It is these modulations which in fact allow us to determine the inelastic spectral density from the optical data.

When the gap is anisotropic the region around $\omega = 2\Delta_0$ gets smeared out and the onset in $1/\tau^{op}(\omega)$ is no longer vertical but rather shows a more gradual increase with its profile related to an average over the variation in gap value as a function of momentum directions. While the details of the variation in gap value with direction are not directly reflected in this quantity, nevertheless our fits show that anisotropy is required in order to be able to understand the data presented in reference \cite{min:2013} in the specific case of LiFeAs at $T =$ 4 K in the superconducting state. While we cannot compare our results directly with the much more detailed information on gap anisotropy provided by ARPES in reference \cite{borisenko:2012,umezawa:2012} both sets of data are consistent in that they show considerable anisotropy.

The recovered electron-boson spectral density found in our work for LiFeAs shows similarities with those found (also from optics) in the high $T_c$ cuprates. All spectra show a distinct resonant peak $\Omega_R$ at low energies which scales roughly like $\Omega_R^{OPT} \cong$ 6.3 $k_B T_c$. This, on its own, could be taken as evidence for a common mechanism between these two classes of materials but, as we argue below, this may not be so. In the case of LiFeAs the resonant peak seen in optics, $\Omega_R^{OPT}$, is equal to $\approx$ 8 meV. This energy is very close to the energy of the prominent structures seen in scanning tunneling data (STS)\cite{chi:2012} in this same material and provides strong evidence for consistency between optics and tunneling. Note that both techniques involve an average over all momenta ($\vec{k}$) in contrast to ARPES which is $\vec{k}$ sensitive. The relationship between the peak in $I^2\chi(\omega)$ seen in optics and spin fluctuations is, however, not so clear. While its energy corresponds nicely to the energy of the peak seen in the incommensurate spin fluctuation spectrum\cite{qureshi:2012} at $\vec{Q}_{inc}$ the intensity of this spectrum is very weak. Further, optics gives no compelling evidence that this peak at 8 meV is suppressed and merges into the background in the normal state as would be expected if it is due to superconductivity\cite{he:2001,he:2002}. This does not eliminate the possibility that the broad peak we get around 8 meV and the extended spectrum to high energies is due, in part or even entirely, to a spin fluctuation background which would then make the main contribution to the pairing glue in this material. As a further note, we have found no clear spectroscopic signature in the $I^2\chi(\omega)$ that we recovered from optics, for the presence of an important phonon contribution from sharp modes at 15, 30 and 44 meV. This is in contrast to an interpretation of ARPES data which concludes that phonons may dominate the inelastic scattering seen in the quasiparticle self energies along two cuts on the hole Fermi surfaces about the $\Gamma$ point in LiFeAs\cite{kordyuk:2011}. This observation may, however, not be inconsistent with our results because optics is not momentum specific and instead involves a momentum average. A point of agreement between the two sets of data worth noting is that both find important coupling to very low energy excitations in LiFeAs and this is different from what is found in the high $T_c$ cuprates.

\ack
We thank E. Schachinger for carrying out the Eliashberg calculations used in this study and Sarah Burke and Bruce Gaulin for useful discussions. JH acknowledges financial support from the National Research Foundation of Korea (NRFK Grant No. 2013R1A2A2A01067629). JPC and TT were supported by the Natural Science and Engineering Research Council of Canada (NSERC) and the Canadian Institute for Advanced Research (CIFAR). YSK was supported by the Basic Science Research Program
(NRF-2013R1A1A2009778) and the Leading Foreign Research Institute Recruitment Program (Grant No. 2012K1A4A3053565).

%
%

\section*{References}
\bibliographystyle{unsrt}
\bibliography{bib}

\end{document}